\documentclass[conference]{IEEEtran}

\pdfoutput=1

\usepackage{microtype}

\usepackage[utf8]{inputenc}

\usepackage[T1]{fontenc}
\usepackage{amsmath}
\usepackage{bm}
\usepackage{xspace}

\usepackage[english]{babel}

\usepackage{hyperref}

\usepackage{todonotes}

\usepackage{siunitx}

\usepackage{subfig}

\usepackage{dblfloatfix}

\usepackage{blindtext}

\usepackage{booktabs}

\usepackage{listings}

\usepackage{color, colortbl}
\definecolor{Gray}{gray}{0.9}
\definecolor{LightGray}{gray}{0.97}

\usepackage{textcomp}

\usepackage{xcolor}

\usepackage{marginnote}

\usepackage{marvosym}

\usepackage{hyperref}

\usepackage[normalem]{ulem}

\usepackage{fancyhdr}
\definecolor{not-supported}{RGB}{250,150,100}
\definecolor{supported}{RGB}{50,250,50}
\definecolor{unknown}{RGB}{250,250,50}

\newcommand{\figref}[1]{Figure~\ref{#1}}
\newcommand{\lstref}[1]{Listing~\ref{#1}}
\newcommand{\tabref}[1]{Table~\ref{#1}}
\newcommand{\secref}[1]{Section~\ref{#1}}

\begin{document}
	
\captionsetup[table]{labelfont={footnotesize,sc}, textfont={footnotesize,sc}}
\title{Energy Efficiency Features of the Intel Skylake-SP Processor and Their Impact on Performance}

\author{\IEEEauthorblockN{Robert Schöne, Thomas Ilsche, Mario Bielert,  Andreas Gocht, Daniel Hackenberg}
	\IEEEauthorblockA{Center for Information Services and High Performance Computing (ZIH)\\
		Technische Universit\"{a}t Dresden -- 01062 Dresden, Germany\\
		Email: \{robert.schoene | thomas.ilsche | mario.bielert | andreas.gocht | daniel.hackenberg\}@tu-dresden.de\\
	}
}

\maketitle

\cfoot{\vspace{-1cm}\footnotesize \textbf{© 2019 IEEE}. Personal use of this material is permitted. Permission from IEEE must be obtained for all other uses, in any current or future media, including reprinting/republishing this material for advertising or promotional purposes, creating new collective works, for resale or redistribution to servers or lists, or reuse of any copyrighted component of this work in other works. The definite version is available at \url{https://doi.org/10.1109/HPCS48598.2019.9188239}.}
\renewcommand{\headrulewidth}{0pt}
\thispagestyle{fancy}
\begin{abstract}
The overwhelming majority of High Performance Computing (HPC) systems and server infrastructure uses Intel x86 processors.
This makes an architectural analysis of these processors relevant for a wide audience of administrators and performance engineers.
In this paper, we describe the effects of hardware controlled energy efficiency features for the Intel Skylake-SP processor.
Due to the prolonged micro-architecture cycles, which extend the previous Tick-Tock scheme by Intel, our findings will also be relevant for succeeding architectures.
The findings of this paper include the following:
C-state latencies increased significantly over the Haswell-EP processor generation.
The mechanism that controls the uncore frequency has a latency of approximately 10\,ms and it is not possible to truly fix the uncore frequency to a specific level.
The out-of-order throttling for workloads using 512 bit wide vectors also occurs at low processor frequencies.
Data has a significant impact on processor power consumption which causes a large error in energy models relying only on instructions.
\end{abstract}
\IEEEpeerreviewmaketitle

\section{Introduction and Background}

\label{sec:intro}
In recent years, the number of energy efficiency features in Intel processors increased significantly.
Contemporary processors support Per-Core P-States (PCPs)~\cite{haswell_ee}, Uncore Frequency Scaling (UFS)~\cite{haswell_ee}, turbo frequencies~\cite[Section 14.3.3]{Intel3}, core and package C-states~\cite{cstate-latency}, T-states~\cite{tstates}, power capping~\cite{rapl_powerbound}, Energy-Performance-Bias (EPB)~\cite[Section 14.3.4]{Intel3}, and other mechanisms.
While these features improved the energy-proportionality significantly, they also have a major influence on the performance of the processor.
This can be seen when comparing the given system configuration for results submitted for the SPECpower\_ssj~\cite{specpower} with those submitted to performance related benchmarks like SPEC CPU2017~\cite{speccpu}.
While the former results got more energy proportional within the last decade, the latter often disable power saving mechanisms to increase performance and provide repeatable results.
Furthermore, new architectural features, which boost performance, also influence the power consumption.
One example is the introduction of AVX2 and FMA in Intel Haswell processors.
Some workloads using these features, would increase the power consumption over the thermal design power (TDP) at nominal frequency.

All of these power-related features span a huge design space with conflicting optimization goals of performance, power capping, and energy efficiency.
Users are often oblivious to the implications of these mechanisms which are typically controlled by hardware, firmware and sometimes the operating system.
Therefore, they are using their systems as is, hoping that the applied settings will fit their purpose.
High Performance Computing (HPC), however, has specific requirements because a single wrongly-configured core in a highly parallel program can lead to millions of cores waiting.
Therefore, a detailed analysis and understanding of the possible impacts of power saving mechanisms is key to every performance and energy efficiency evaluation and optimization.

In this paper, we describe new features of Intel Skylake-SP processors, which are likely also part of future processors.
In previous years, Intel followed a Tick-Tock scheme, where a new architecture and a new process are introduced alternately.
Recently, this cycle has been extended with a third step where the architecture is slightly improved.
This is a chance for software and performance engineers, because
an adaption of software for a specific hardware has now more time to be worthwhile.

This paper is structured as follows:
\secref{sec:architecture} describes relevant micro-architectural changes and energy efficiency features of the Intel Skylake-SP processor compared to the predecessor generations.
In \secref{sec:test_system_setup}, we give an overview about the test system that we use for our experiments.
A short analysis of standardized mechanisms described in the Advanced Configuration and Power Interface (ACPI) is given in \secref{sec:acpi-states}.
Sections~\ref{sec:uncore-freq}-\ref{sec:power} present an analysis of hardware controlled energy efficiency mechanisms and their effect on runtime and power consumption.
We conclude our paper with a summary and an outlook in \secref{sec:summary}.

\section{Skylake-SP Architecture and Energy Efficiency Features}
\label{sec:architecture}
The Skylake server processors introduce several new micro-architectural features, which increase performance, scalability, and efficiency.
Furthermore, they uphold energy efficiency features from older architectures and introduce new ones.

\subsection{Microarchitecture}

On the core level, which is summarized in \tabref{tab:skl_arch}, AVX-512 (or AVX512F to be precise) is the most anticipated feature for HPC.
With AVX-512, an entire cache line (64\,B, one ZMM register) can now be used as input for vector instructions.
Furthermore, with the new \texttt{EVEX} prefix, the number of addressable SIMD registers doubled to 32.
Therefore, 2\,kiB of data can now be held in vector registers, compared to 512\,B in Haswell and Broadwell architectures.
The L1D cache bandwidth has been doubled while L2 cache bandwidth is still limited to 64\,B/cycle.
The L2 cache size increased from 256\,kiB to 1\,MiB, whereas LLC slice sizes decreased from 2.5\,MiB to 1.375\,MiB.
The increased number of store buffers can lead to a faster streamed store throughput.
In addition, all codes benefit from extended out-of-order features.

\begin{table}[t!]
\centering
\caption{\label{tab:skl_arch}Comparison of Haswell-EP and Skylake-SP processors}
\begin{tabular}{l l l} \hline
\toprule
\textbf{Microarchitecture}	& \textbf{Haswell-EP}			& \textbf{Skylake-SP 6100/8100}		\\
\midrule
References & \cite{agner,Intel_HSW,Intel_HSW_ep,intel_optimization} & \cite{agner,intel_optimization,intel_skylake_sp_isscc,skylake_overview} \\
Allocation queue	&  56/core	& 64/thread		\\ 
\rowcolor{LightGray}Execute			& 8\,micro-ops/cycle & 8\,micro-ops/cycle	\\
\rowcolor{LightGray}Scheduler entries	& 60					& 97			\\
ROB entries		& 192					& 224			\\
\rowcolor{LightGray}INT/FP register file	& 168/168				& 180/168		\\
SIMD ISA		& AVX2					& AVX-512			\\ %
\rowcolor{LightGray}FPU width		& 2\texttimes 256 Bit FMA	& 2\texttimes512 Bit FMA	\\ 
FLOPS/cycle\,(double)	& 16					& 32			\\
\rowcolor{LightGray}Load/store buffers	& 72/42					& 72/56			\\ %
L1D accesses	  	& 2\texttimes 32\,B load +			& 2\texttimes 64\,B load +	\\
per cycle               & 1\texttimes 32\,B store			& 1\texttimes 64\,B store	\\
\rowcolor{LightGray}L2 B/cycle		& 64					& 64			\\
Supported memory 	& 4\texttimes DDR4-2133				& 6\texttimes DDR4-2666 		\\
DRAM bandwidth		& up to 68.2\,GB/s			&  up to 128 GB/s	\\
\bottomrule
\end{tabular}
\end{table}

\begin{figure}[b]
\center
\includegraphics[width=\columnwidth]{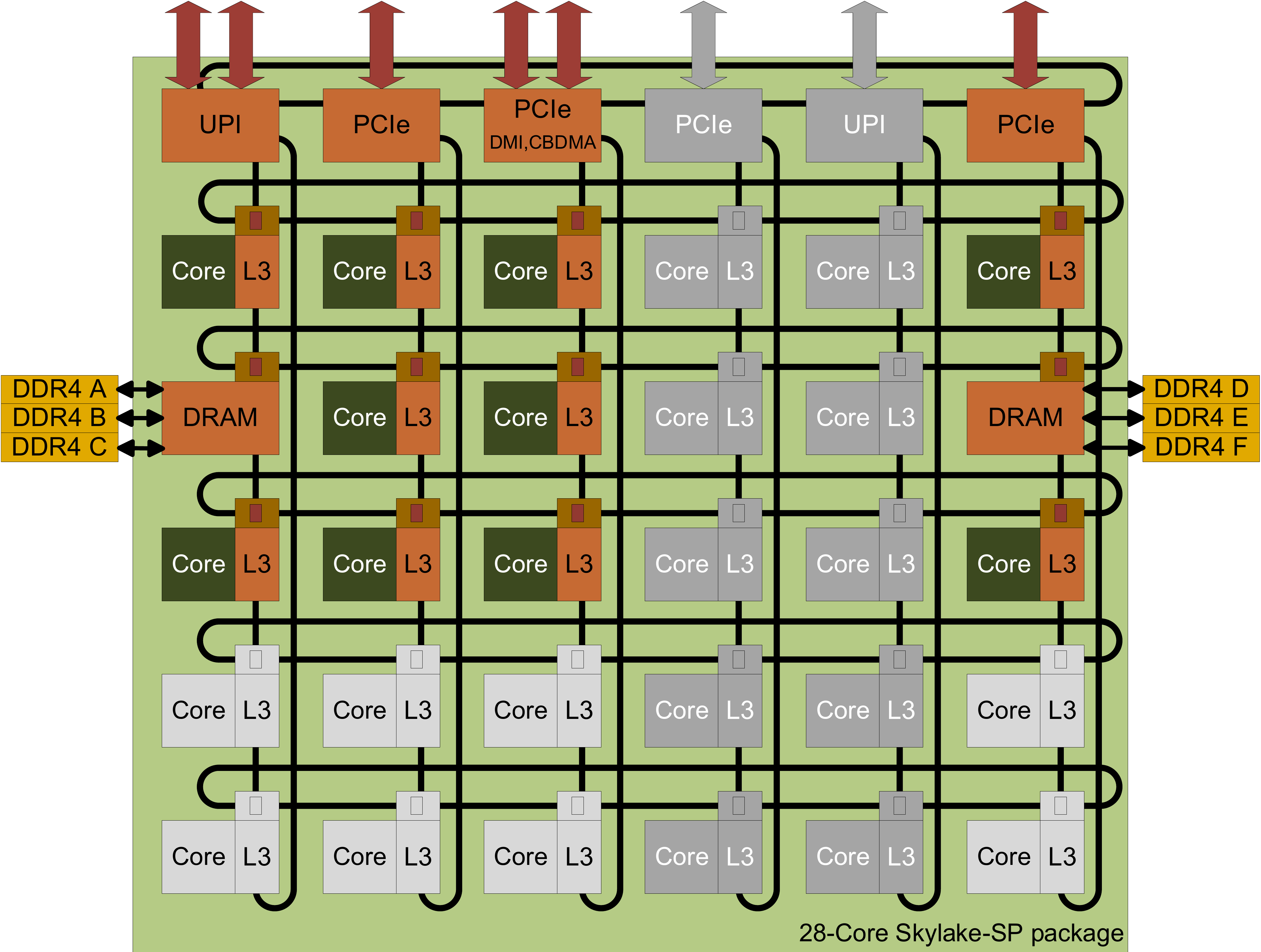}
\caption{\label{fig:skl_package} 28 core package Skylake-SP, dark gray components are removed in 18 core processors, light and dark gray components are removed for 10 core packages}
\end{figure}

The uncore also faced a major re-design.
In previous architectures (starting with Nehalem-EX), cores have been connected by a ring network.
The increased number of cores in Haswell and Broadwell architectures led to the introduction of a second ring with connections between both.
Skylake now introduces a mesh architecture, where each core, including its LLC slice, connects to a 2D mesh, as depicted in \figref{fig:skl_package}.
The external communication (PCIe, UPI) is placed on one side of the mesh.
Furthermore, two cores in the remaining mesh are replaced by integrated memory controllers (iMCs), each of which can host up to three DRAM channels.
There are multiple flavors of Skylake-SP based processors with varying external connections and numbers of cores.

The cluster-on-die (CoD) feature is now called sub-NUMA clustering and enables dividing the cores into two NUMA domains.
Cache coherence is implemented by MESIF and a directory-based home snoop protocol.

\subsection{Energy Efficiency Mechanisms}
Like its predecessors, Skylake-SP processors support Per-Core P-States (PCPs) and Uncore Frequency Scaling (UFS).
This enables fine-grained control over performance and energy efficiency decisions.
The Energy Performance Bias (EPB) indicates, whether to balance the profile for runtime or power consumption or something in between.
The Energy-Efficient Turbo (EET) mechanism was already available on the Haswell platform.
All these features have been described in detail in~\cite{haswell_ee}.

The Turbo Boost Max 3.0 (TBM3) feature was introduced with Broadwell-E processors.
Its basic idea is to improve single thread performance by executing the workload on the processor core that delivers the best power and performance.
Due to variations in their manufacturing process, single processors and cores can have varying efficiency characteristics.
This facilitates higher turbo frequencies on some cores and lower on others, which in turn influences the performance of a single thread depending on the hardware core that executes it.
However, currently none of the currently available Xeon Scalable Processors supports this feature.
An anticipated feature of Skylake processors is Hardware Duty Cycling (HDC)~\cite[Section 14.5]{Intel3}, also known as SoC Duty Cycling.
HDC implements a more coarse grained duty cycling in comparison to T-states.
In contrast to clock gating, HDC uses C-states with power gating.
However, this feature is targeted at mobile and desktop processors and is not described in more detail in this document.

\subsection{Hardware-Controlled P-States (HWP)}
\label{sec:theory:hwp}
Introduced with the Broadwell generation of Intel processors, Hardware-Controlled P-States (alias Hardware Power Management (HWPM) or SpeedShift), move the decision of frequency and C-state usage from the operating system to the processor~\cite[Section 14.4]{Intel3}.
This removes the perturbation of an OS control loop, which interrupts the workload regularly.
Furthermore, it increases the responsiveness because the hardware control loop can be executed more frequently without perturbation.
While the Broadwell processors hardware acts mostly autonomously, Skylake-SP processors provide interfaces for a collaboration with the OS through interrupts~\cite[Section 14.4.6]{Intel3}.
With the HWP interface, the OS can define a performance and power profile, and set a minimal, efficient, and maximal frequency.
Under Linux, the tool \texttt{x86\_energy\_policy} can be used to interact directly with the hardware interface.
HWP is accompanied by an incremental counter register \texttt{MSR\_PPERF}, which holds the Productive Performance Count (PCNT) that should increase only if a cycle has been used efficiently.

\subsection{AVX-512 Frequency}
Current Intel processors are unable to use all of their components concurrently at reference frequency.
With the doubling of the SIMD width to 512 bit, the power consumption of SIMD execution units now varies even more.
This relates to both their static and dynamic power consumption.
The former is reduced by disabling those components of the execution units that calculate upper parts of the register.
Therefore, the possible dynamic power consumption will increase with (a) the width of the vector, (b) the shape of the processed data, (c) the complexity of the operation, and (d) the frequency and voltage that the execution unit uses.
Still, the average power consumption has to stay within the given TDP limit in the long-term.
Since the processor core will use the full SIMD width given by the program and cannot change the processed data or operation, the only remaining option for lowering power consumption is dynamic voltage and frequency scaling (DVFS) via PCPs and UFS.
This leads to three different frequency ranges that have to be considered: \textit{normal frequencies}, \textit{AVX frequencies}, and \textit{AVX-512 frequencies}.
Each ranges from a guaranteed reference to a possible turbo frequency, which depends on the number of active cores.
However, these names can be misleading, since AVX instructions can be distinguished into \textit{heavy} and \textit{light} instructions.
For example, AVX frequencies are applied to workloads that use floating-point (FP) and integer multiplication with 256 bit vector width, but also workloads that use light AVX-512 instructions (i.e., without FP and INT MUL).
Likewise, AVX-512 frequencies are used for workloads with heavy AVX-512 instructions.

However, frequency changes are not processed instantaneously as Mazouz et al describe in~\cite{ftalat}.
Therefore, a different mechanism has to be implemented to lower the power consumption before the frequency is changed.
According to Agner~\cite{agner}, initial AVX or AVX-512 instructions will be issued slower if they have not been used recently.

\section{Test System Setup}
\label{sec:test_system_setup}
For our analysis we use a system equipped with two 18-core Intel Xeon-SP 6154 Gold processors with HyperThreading support running Ubuntu 18.04.1 with a Linux 4.17.0-3 kernel.
An overview on the system is given in \tabref{tab:testsystem}.

With HWP enabled, even memory-bound workload run at the maximum allowed frequency, which is not what we would expect.
According to our measurements, the PCNT register \texttt{MSR\_PPERF} (see \secref{sec:theory:hwp}) increases with every cycle even for workload that are stalled most of the time (e.g., pointer-chasing or streamed data access in DRAM).
We therefore did not use HWP.
EET also does not seem to influence compute-bound/memory-bound workloads and therefore left at the default setting (enabled).
We use the \texttt{acpi\_cpufreq} driver and the \texttt{userspace} governor to control core frequencies and the \texttt{msr} kernel module to change uncore frequencies.
The only available C-states on our system are C1 (HLT), C1E (HLT+DVFS) and C6 (power gating).
The power limits given by RAPL are 240\,W over a period of 1\,second and 200\,W over a period of 100\,seconds.
\begin{table}[t!]
	\centering
	\caption{Test system details}
	\begin{tabular}{ll} \hline
		\toprule
		\rowcolor{LightGray} Processor                                       & 2x Intel Xeon SP 6154 Gold\\
		Frequency range (selectable P-states)   & 1.2 -- 3.0\,GHz\\
		\rowcolor{LightGray} Turbo frequency                                 & up to 3.7\,GHz \\
		AVX frequency range              & 2.6--3.6\,GHz \\
		\rowcolor{LightGray} AVX-512 frequency range  & 2.1--3.5\,GHz \\
		Energy perf. bias  & balanced \\
		\rowcolor{LightGray}Energy-efficient turbo (EET)    & enabled \\
		Uncore frequency scaling (UFS)                  & 1.2-2.4\,Ghz \\
		\rowcolor{LightGray}Hardware Performance States (HWP)                  & disabled \\
		Per-core p-states (PCPs)         & enabled \\
		\midrule
		\rowcolor{LightGray}RAM        		   & 12x 32GB DDR4-2666 \\
		\rowcolor{LightGray}Motherboard        & Tyan S7106 \\
		\midrule
		\rowcolor{LightGray}Idle power (fan speed set to maximum)           & 78\,W \\
		Power meter                      & ZES LMG 450\\
		\rowcolor{LightGray}Accuracy                                        & 0.07\,\% + 0.5\,W \\ 
		\bottomrule
	\end{tabular}
	\label{tab:testsystem}
\end{table}

\begin{figure}[b!]
\center
\includegraphics[width=\columnwidth]{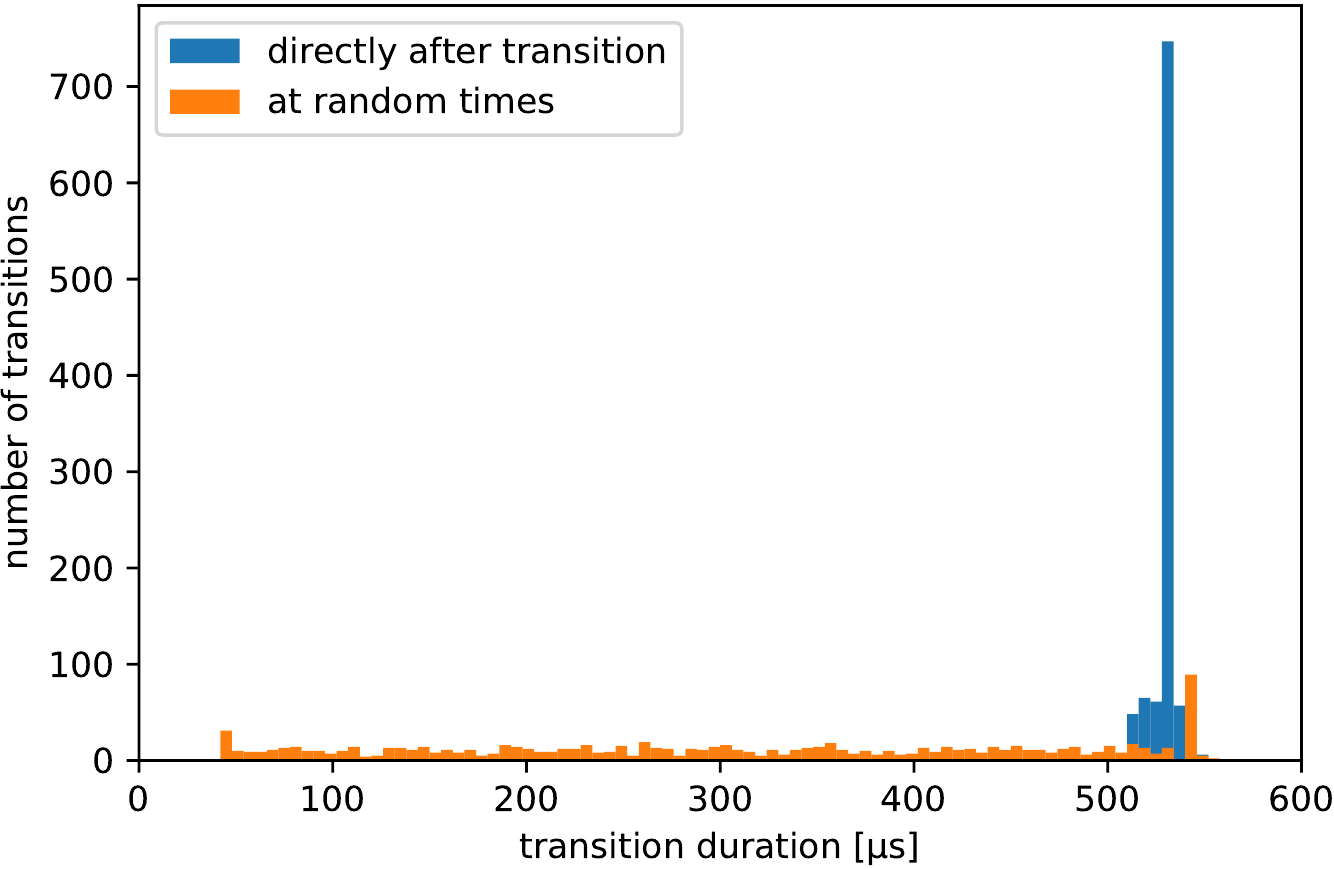}
\caption{\label{fig:ftalat} Exemplary P-state transition time histogram (1.5$\rightarrow$2.6\,GHz). The frequency transition is initiated either at a random time or directly after our PMC measurement detects that the previous frequency change was applied. The distribution for the latter case shows that the 500\,µs update interval applies for Skylake-SP as well.}
\end{figure}

\begin{figure*}[b]
	\subfloat[\label{fig:cstate-local} Local]
	{
		\includegraphics[width=\columnwidth]{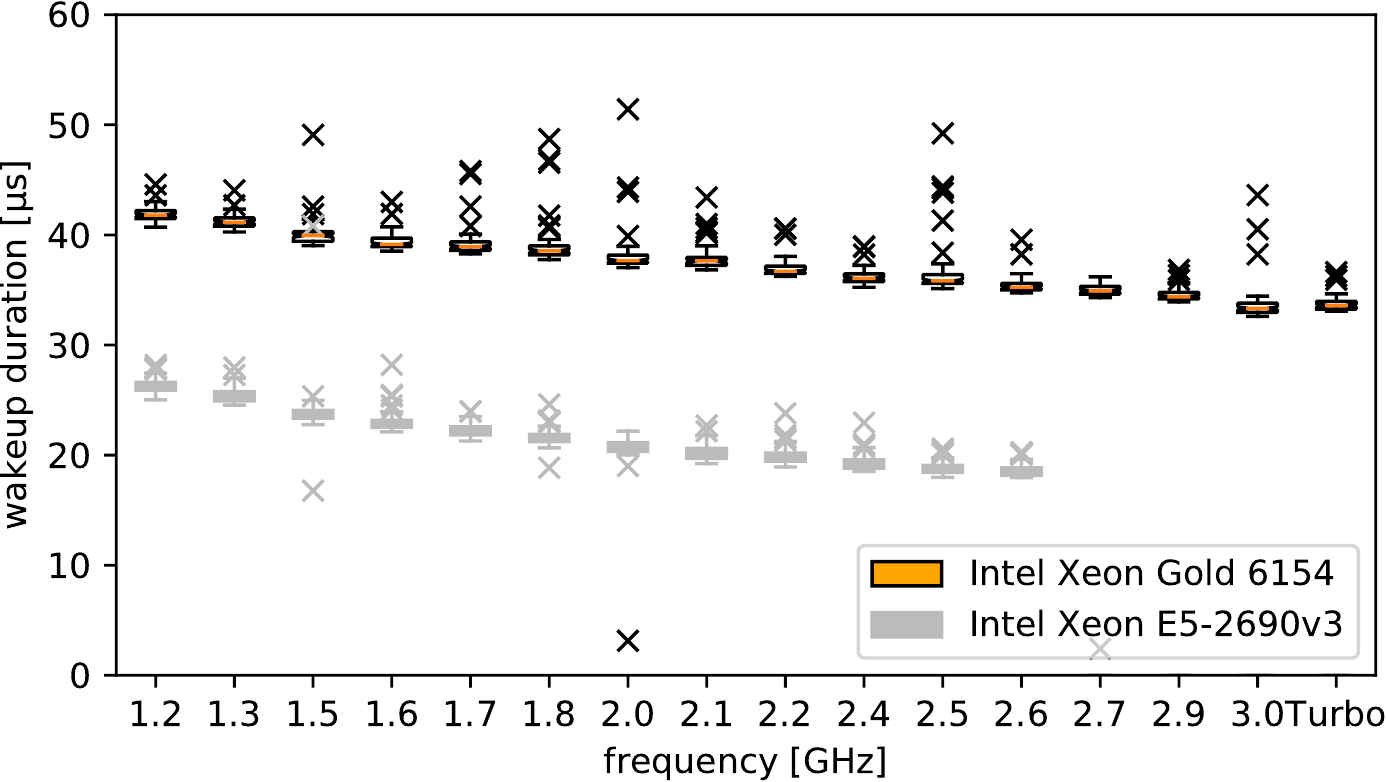}
	}
	\hfill
	\subfloat[\label{fig:cstate-remoteidle} Remote idle]
	{
		\includegraphics[width=\columnwidth]{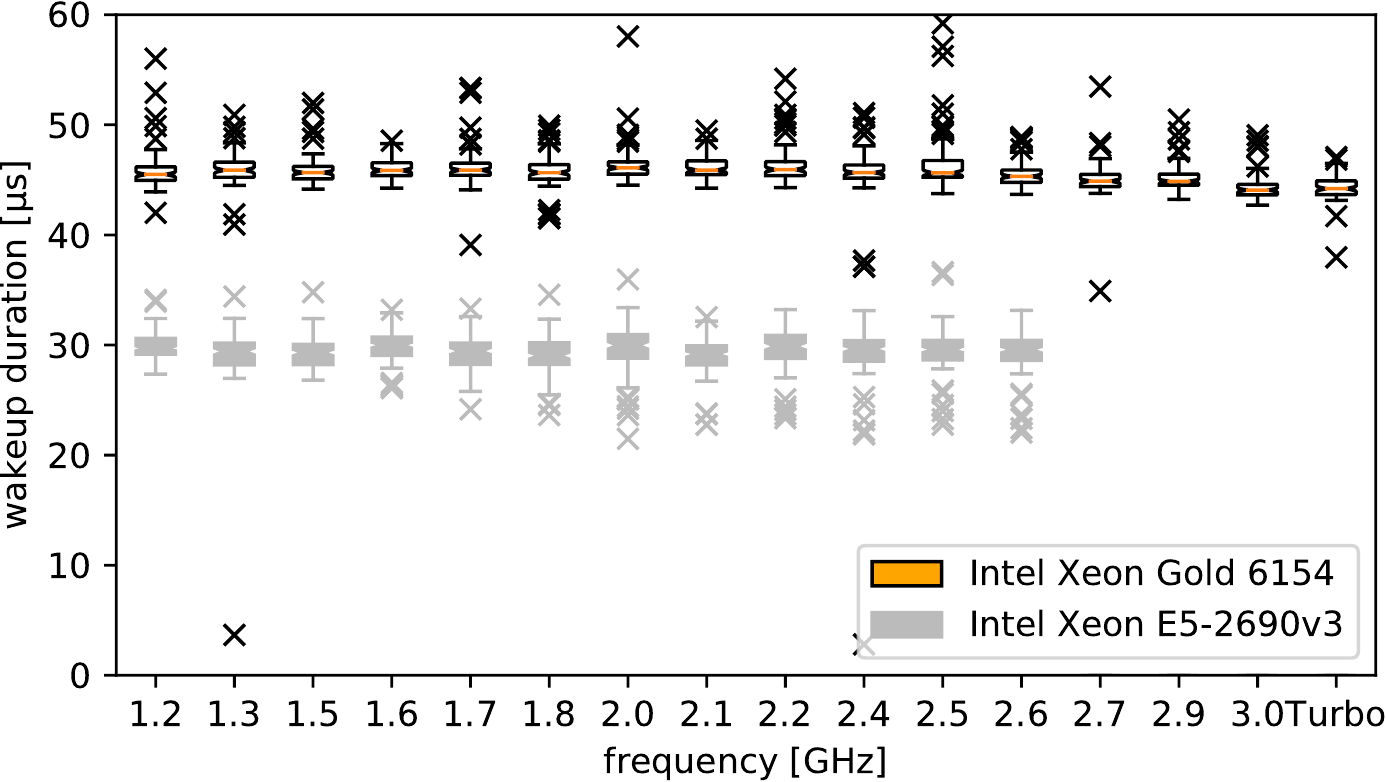}
	}
	\caption{\label{fig:cstate} C-state information for Skylake-SP processor in comparison to Haswell-EP (Intel Xeon E5-2690 v3, gray). Outliers, which imply a fast transition can be explained with a difference between the requested C-state and the actual C-state, which is transparently selected by a hardware mechanism.}
\end{figure*}

\section{Transition Latencies and ACPI State Behavior}
\label{sec:acpi-states}
Traditional energy efficiency mechanisms are defined in the Advanced Configuration and Power Interface (ACPI)~\cite{acpi}.
This includes sets of P-, C-, and T-states, which can be used to lower power consumption.

\subsection{Performance States/P-States/PCPs}
One of the traditional energy saving strategies is to reduce the core frequency of processors.
Previous work illustrates that an external mechanism is responsible for setting core frequencies in Haswell-EP processors~\cite{haswell_ee}.
The mechanism applies changes to the P-state at a regular interval.
The duration of a frequency change depends on the timing of the change request in comparison to the mechanism's internal update interval.
The Skylake-SP processor shows a similar behavior (see \figref{fig:ftalat}).

\subsection{Power States/C-States}
Intel changed several details in their C-state implementation.
Out of five available C-states on Haswell systems, only four remain on Skylake-SP processors since C3 is not defined in the kernel sources. %
The remaining core C-states are C0, C1, and C6.
Additionally, the operating system can request C1E.
Our measurement methodology reuses a previous work from~\cite{cstate-latency} and~\cite{hppac}.
One CPU (\textit{caller}) wakes another CPU (\textit{callee}).
We define the C-state transition latency as the time between initiating the wake-up at caller and the callee becoming active.
To measure the start timestamp, we log the kernel tracepoint \texttt{sched:sched\_wake\_idle\_without\_ipi} on the caller CPU, which is closest to the wake-up event generated in the function \texttt{set\_nr\_and\_not\_polling}.
The return to an active state is monitored using the \texttt{power:cpu\_idle} event on the callee CPU.
To generate the wake-up events, we use a thread signal, as shown in \lstref{lst:cstate-trigger}.
To increase the accuracy, the measurement is repeated 100 times for each combination of C-state and P-state.

\begin{lstlisting}[label=lst:cstate-trigger,caption=\texttt{cond\_wait} program to trigger C-state wake-up, language=C,frame=single,captionpos=b,tabsize=2,basicstyle=\footnotesize]
*callee_work*:
 for (i=0;i<ntimes;++i) {
  pthread_mutex_lock(&lock);
  pthread_cond_wait(&cv, &lock);
  pthread_mutex_unlock(&lock);
 }

*caller_work*:
 for (i=0;i<ntimes;++i) {
  sleep(1);
  pthread_cond_signal(&cv);
 }
\end{lstlisting}

\figref{fig:cstate} shows that a local wake-up call to a core in C6 results in a 42\,µs latency for the lowest core frequency and about 33\,µs for the nominal frequency.
The same is true for a remote active state (not depicted).
If the second socket is idling, the wake-up latency is mostly between 46 and 48 µs.
However, sometimes the latency reaches 55\,µs.
Compared to the Haswell EP platform, the latencies increased significantly.

\subsection{Throttling States/T-States}
Like their predecessors, Intel Skylake processors also support clock modulation as a mechanism to lower power consumption and to prevent thermal damage.
The interface to control these mechanisms through software remained unchanged.
Using a previously established methodology~\cite{tstates} we conclude that the performance characteristics are similar to the Haswell-EP architecture~\cite{haswell_ee}:
Each core independently controls its own T-state.
In contrast to the Sandy Bridge architecture, no DVFS is applied.
The highest T-state (with the most cycles being skipped) is not implemented.
The lower T-states skips more cycles than defined.

\section{Uncore Frequency Scaling/UFS}
\label{sec:uncore-freq}

\setcounter{figure}{4}
\begin{figure*}[b]
	\subfloat[\label{fig:ufs-tgaphist} Histogram $t_{gap}$ (1.4$\rightarrow$2.4 GHz)]
	{
		\includegraphics[width=.3\textwidth]{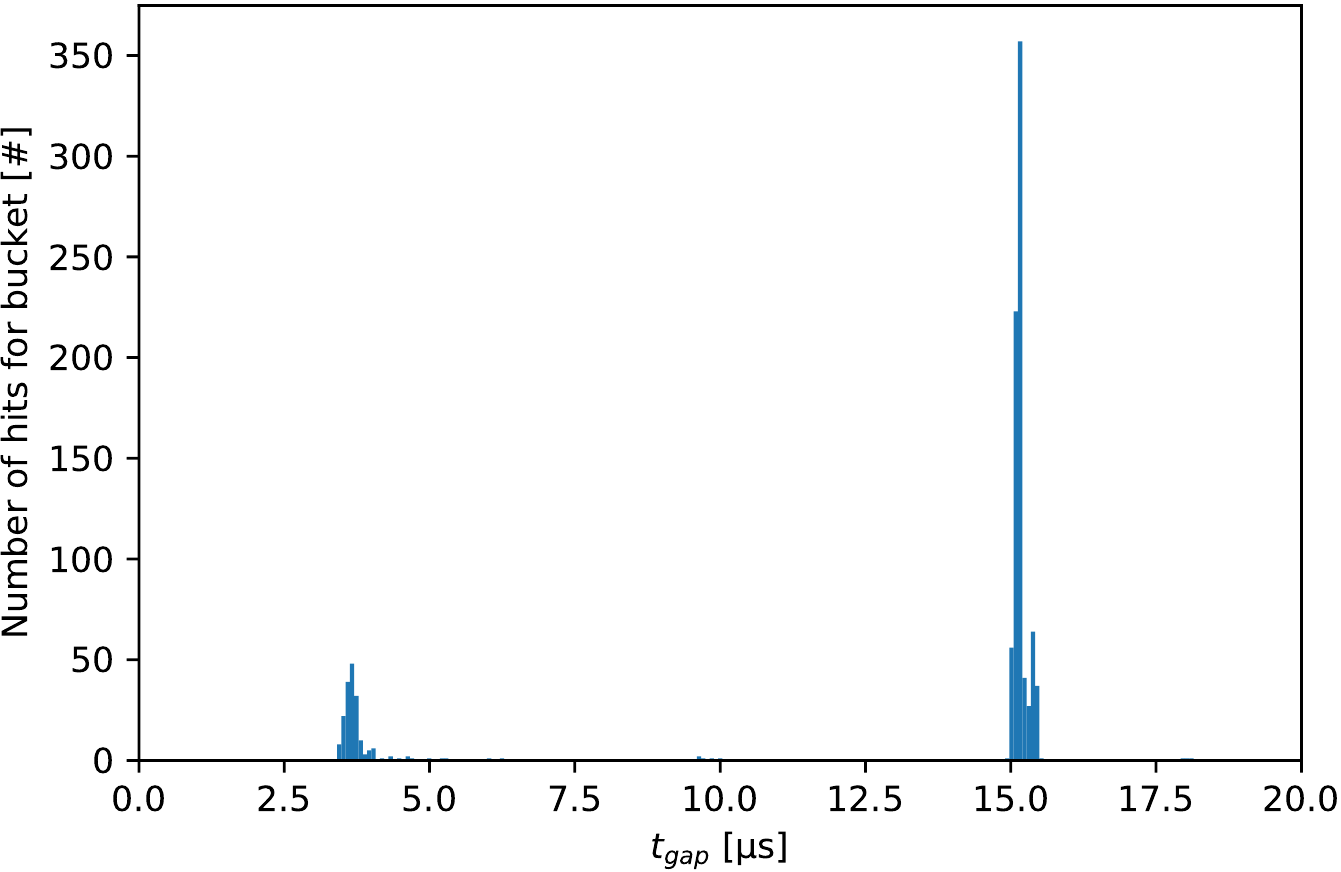}
	}
	\hfill
	\hfill
	\subfloat[\label{fig:ufs-tdelay} Histogram $t_{delay}$ (filtered, 1.4$\rightarrow$2.4 GHz)]
	{
		\includegraphics[width=.3\textwidth]{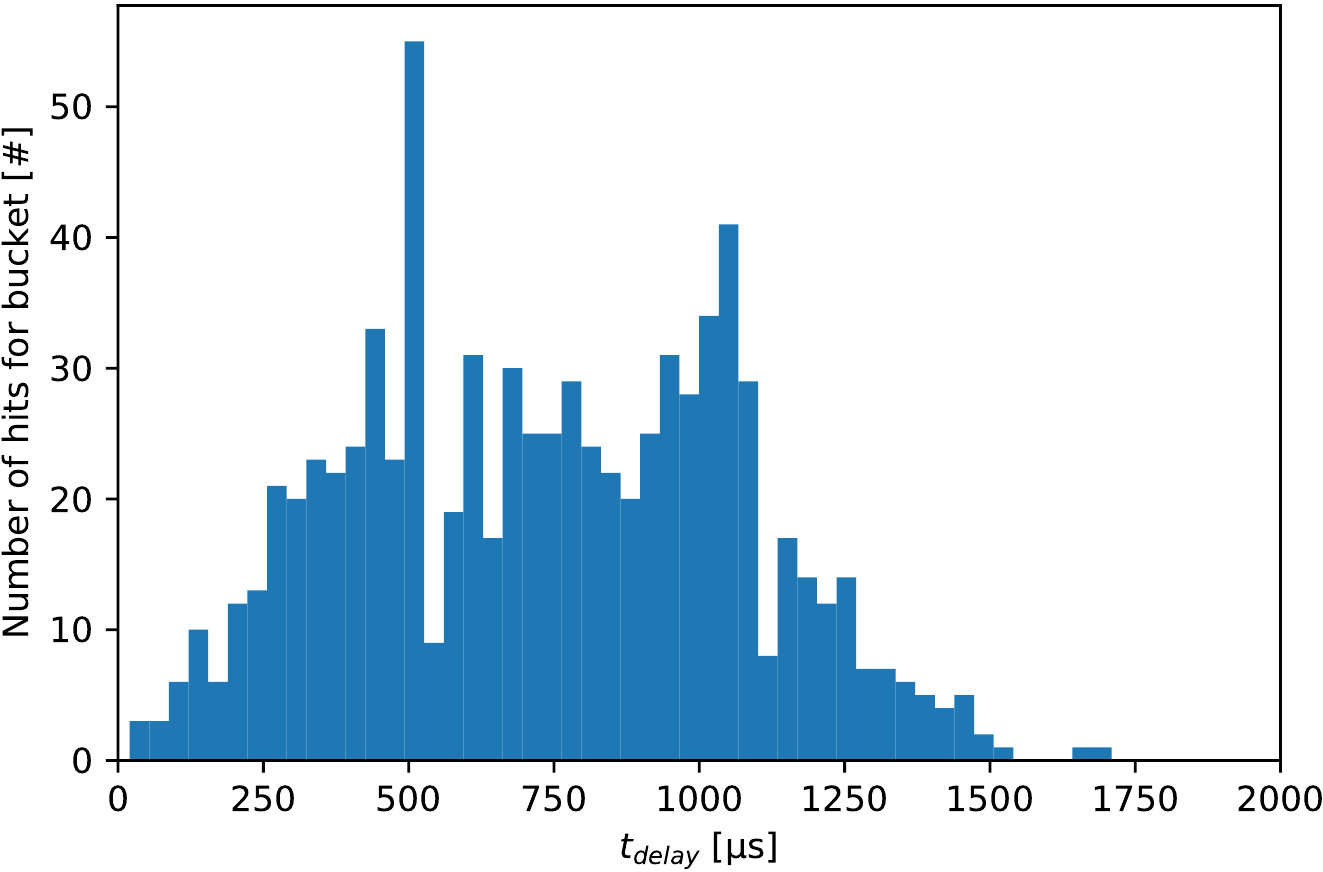}
	}
	\hfill
	\subfloat[\label{fig:ufs-tinitialize} Histogram $t_{controlloop}+t_{delay}$]
	{
		\includegraphics[width=.3\textwidth]{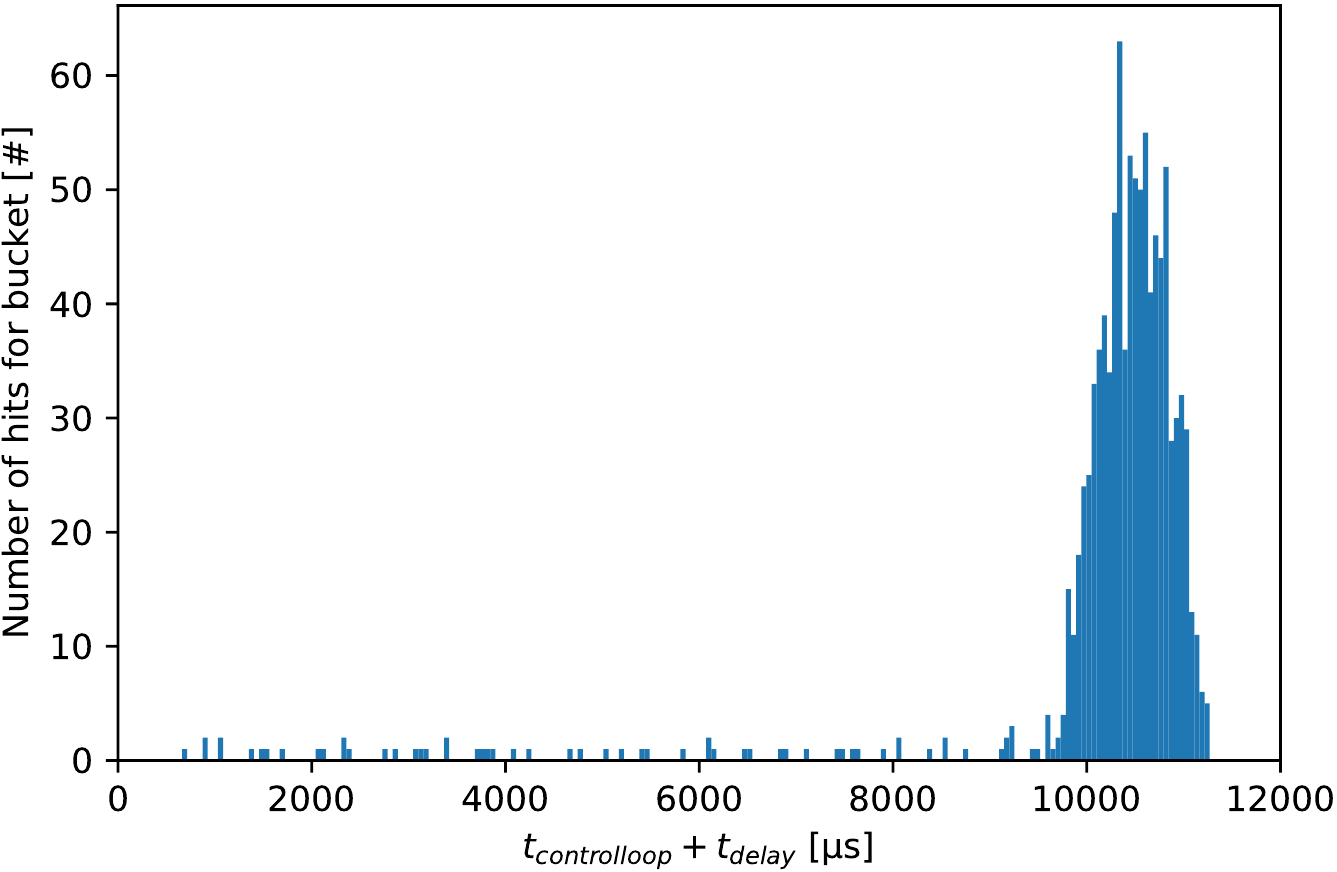}
	}
	\caption{\label{fig:ufs} Uncore frequency measurement results}
\end{figure*}

\setcounter{figure}{3}
\begin{figure}[t]
\center
	\includegraphics[width=.95\columnwidth]{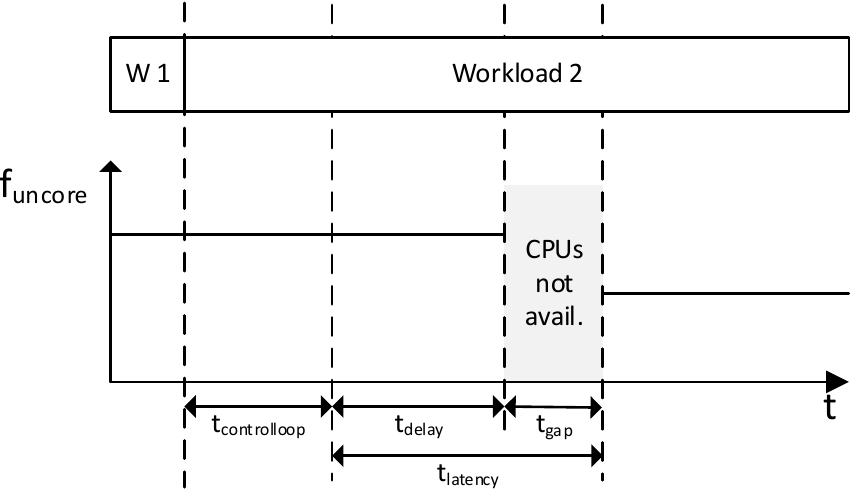}
	\caption{\label{fig:uncore-meas} Important latencies for uncore frequency switches -- $t_{controlloop}$: The time the internal mechanism needs to detect the changed workload; $t_{delay}$: The time until the frequency transition starts; $t_{gap}$: The time the processor is not available.}
\end{figure}

\setcounter{figure}{5}
The UFS feature enables processors to change the frequency of non-core components during runtime independently from core frequencies~\cite{haswell_ee}.
By default, the uncore frequency is guided by a hardware control loop.
It can also be controlled through an interface via the model specific register (MSR) \texttt{UNCORE\_RATIO\_LIMIT}~\cite{ee_servers}.
After booting, this register holds descriptions of the minimal and maximal uncore frequency available.
Writing to the register during runtime can narrow the configuration space to a subset or only one frequency setting by providing the same value for both entries.

We investigate several performance characteristics as depicted in~\figref{fig:uncore-meas}:
First, the latency $t_{delay}$ between the initiation and the execution of a frequency change.
Second, the period $t_{gap}$ in which the processor is unavailable due to switching voltage and frequency.
Third, the latency $t_{controlloop}$ until the internal scaling mechanism recognizes a change of the workload.

We extend the \texttt{FTaLaT} methodology to determine the transition latency~\cite{ftalat}.
Originally, the performance of a CPU-bound workload is measured after a processor core frequency change until it reflects the performance expected for the target frequency.
In our setup, we use a pointer-chasing algorithm that accesses data in the last level cache (LLC) and read the time stamp counter (TSC) after every access.
As soon as one iteration took more than 20,000 cycles, we assume a frequency change.
We measure $t_{delay}$ and $t_{gap}$ using the TSC information and calculate the average access latency to LLC before and after the frequency switch.

To measure $t_{controlloop}$ we run the pointer-chasing algorithm that accesses cache lines from level 1 cache multiple times to train the CPU towards low uncore usage and therefore low uncore frequency.
We then run this algorithm within an LLC data set and measure the same performance data as before.
We assume that running the algorithm within the LLC leads to a higher frequency setting and therefore requires an uncore frequency transition.
Finally, we use the reported average latencies to verify that the assumed low and high frequencies have been applied.
All workloads are executed at a 2.4\,GHz core frequency.
\figref{fig:ufs} presents the results.

\figref{fig:ufs-tgaphist} suggest that about 20\,\% of the samples for each source/target frequency pair will have a short $t_{gap}$ of 3 to 4\,µs.
However, these 20\,\% also correlate with outliers in the reached target performance in terms of cycles per access.
For these samples, the gap detection did not work.
They are therefor considered outliers and are not used in the further analysis. %
The remaining results for $t_{gap}$ are in the range of 14.5 to 16\,µs with no clear pattern for different source and target frequencies.
$t_{delay}$ shows a wide range of measured values, which range from 0 to 1.5\,ms (\figref{fig:ufs-tdelay}).
We assume that the same mechanism as for P-state changes applies, but the resolution of the uncore mechanism is about 3 times lower.
\figref{fig:ufs-tinitialize} shows that the default control loop for setting the uncore frequency adds another 9.8\,ms to the delay.
Therefore, the uncore frequency will only be adapted 10\,ms after a changed workload pattern, which can have a significant impact on regularly altering workloads.
According to the measured access times before and after the frequency switch, the control loop changed the uncore frequency from 1.4 to 2.4\,GHz, which corresponds to access times of 119 and 83 cycles, respectively.
A possible optimization to prevent the uncore from clocking down without MSR access would be to use the second hardware threads to access the LLC.

\section{Power Limiting Mechanisms}
\label{sec:firestarter}

The impact of power limiting can be analyzed by maximizing power consumption of the test system.
FIRESTARTER~\cite{firestarter} is a full-system stress test, which targets a high power consumption for memory, processor, and GPU accelerators.
To adapt to the Skylake architecture we introduced AVX-512 instructions to FIRESTARTER, added 512-bit wide broadcast instructions, and changed the memory access parameters.
We use Linpack and the mprime stress test\footnote{Intel implementation and compiler 18.0.2, 65k elements (Linpack); mprime 29.4 build 7; FIRESTARTER 1.7.3} as reference for comparison.
For all three workloads we compare the core cycle rate and AC power consumption as well as their variation over the full execution, all of which was recorded with \texttt{lo2s}~\cite{lo2s}.
Turbo frequency is enabled, all remaining parameters set to their default values.
According to the RAPL interface all workloads hit the maximal TDP of 200\,W per processor.
However, the dynamic behavior differs between the three workloads.

Linpack consists of alternating phases of computation and synchronization~\cite{firestarter}.
The stable peak power consumption is only reached during computation phases, whereas power consumption during synchronization is reduced to $<$430\,W.
Whenever the computational phase starts, there is a $\sim$1\,s spike of $\sim$665\,W, using the turbo headroom from the previous underutilization.
During computational phases, power consumption averages 620\,W with some noise between 573\,W and 664\,W at the measured 50\,ms intervals.
The effective core frequencies vary between threads and time within the computational phases from 2.6\,GHz to 3.2\,GHz, while being stable at 3.7\,GHz throughout the synchronization phases.

Tests with mprime use in-place large FFTs, which are the most power consuming on our system.
mprime executes different specific FFT functions, that slightly differ in power consumption and core frequency.
Overall power varies between 619\,W and 645\,W with an average of 634\,W and core frequency varies from 2.7\,GHz to 3.1\,GHz, both mostly depending on the currently executed functions.

FIRESTARTER runs a constant architecture-specific workload resulting in a high power consumption and low core frequency.
There is a trend of increase in power consumption from 625\,W to 629\,W for the first 7 minutes, likely caused by the temperature leveling out.
Later, power remains between 627\,W and 631\,W, while the frequency differs between the two packages at 2.55\,GHz / 2.67\,GHz.
FIRESTARTER does not reach the average power level of mprime, but triggers a significantly lower frequency.
As future work, FIRESTARTER should be extended to make use of both AVX-512 and 256 bit wide AVX in order to use port 0 and 1 to compute 256 bit per cycle.
Energy efficiency mechanisms have a significant effect on processor stress tests.
FIRESTARTER utilizes the full package power for core frequencies at or above 2.4\,GHz.
Depending on the core frequency, available budgets are shifted to the uncore.
Mprime reaches the given TDP at higher frequencies, typically around the nominal of 3.0\,GHz.
Even if the uncore frequency is fixed by using the MSR 0x620 and the specifications would allow lowering the core frequencies instead: the power limit is reached, the processor lowers the uncore clock.
Therefore, the specified uncore frequency must be regarded as a recommendation under power constrained circumstances.
On Haswell-EP processors, \textit{disabling the turbo mechanism} during power-constrained workloads may \textit{increase} core frequencies~\cite[Table IV]{haswell_ee}.
According to our measurements on the Skylake-SP platform, this effect emerges now when requesting a frequency that is \textit{below the nominal frequency} (e.g., 2.9\,GHz) and cannot be observed if the EPB is set to \textit{performance}.
While the impact of the frequency increase (in this case 20\,MHz) is limited, the effect itself is counter-intuitive.
Therefore, it serves as an example for complex interactions within the microarchitecture that have unexpected side effects.
\section{Performance Impact of AVX Frequency Transitions}

As discussed in \secref{sec:architecture}, Intel Skylake-SP processors operate on three frequency ranges: standard, AVX and AVX-512.
Switching between these is hardware-controlled and influences the performance of the executed workload:
\begin{itemize}
	\item When switching from one category of instructions (i.e., normal or AVX) to more demanding instructions (i.e., AVX or AVX-512), the out-of-order (OoO) engine is throttled to prevent thermal damage during frequency transition.
	\item When switching from one category of instructions (i.e., AVX, AVX-512) to less demanding instructions, the processor remains within the reduced frequency range for a short period of time to avoid unnecessary frequency changes. %
\end{itemize}

\begin{figure}[b]
\center
	\includegraphics[width=.95\columnwidth]{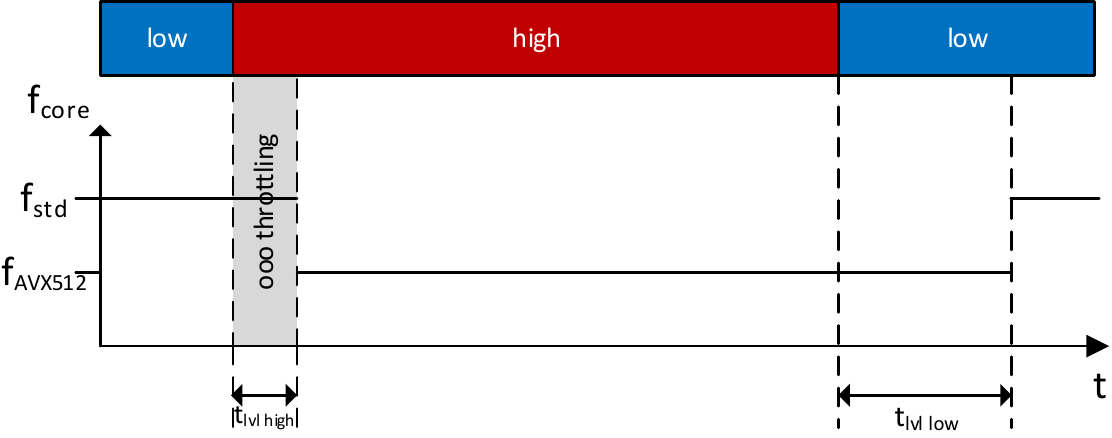}
	\caption{\label{fig:AVX-measurement} AVX frequency transitions}
\end{figure}

To gather information about these state changes, we use a modified version of FIRESTARTER that repetitively switches between \textit{High} (power) and \textit{Low} (power) workloads in defined intervals.
\textit{High} executes the standard workload of FIRESTARTER.
In contrast, the \textit{Low} phase utilizes long running instructions like \texttt{mfence} and \texttt{cpuid}.
\figref{fig:AVX-measurement} illustrates these transitions as $t_{\texttt{lvl\,high}}$ and $t_{\texttt{lvl\,low}}$, respectively.
A manual \mbox{Score-P}~\cite{scorep} instrumentation exposes changes between \textit{Low} and \textit{High} and allows us to record two additional counters:

\begin{itemize}
	\item The \texttt{CORE\_POWER.THROTTLE} event represents the cycles where the out-of-order engine is throttled.
	\item The \texttt{CORE\_POWER.LVL2\_TURBO\_LICENSE} event represents the cycles spent in AVX-512 frequencies.
\end{itemize}

\setcounter{figure}{7}
\begin{figure*}[b]
\center
	\includegraphics[width=\textwidth]{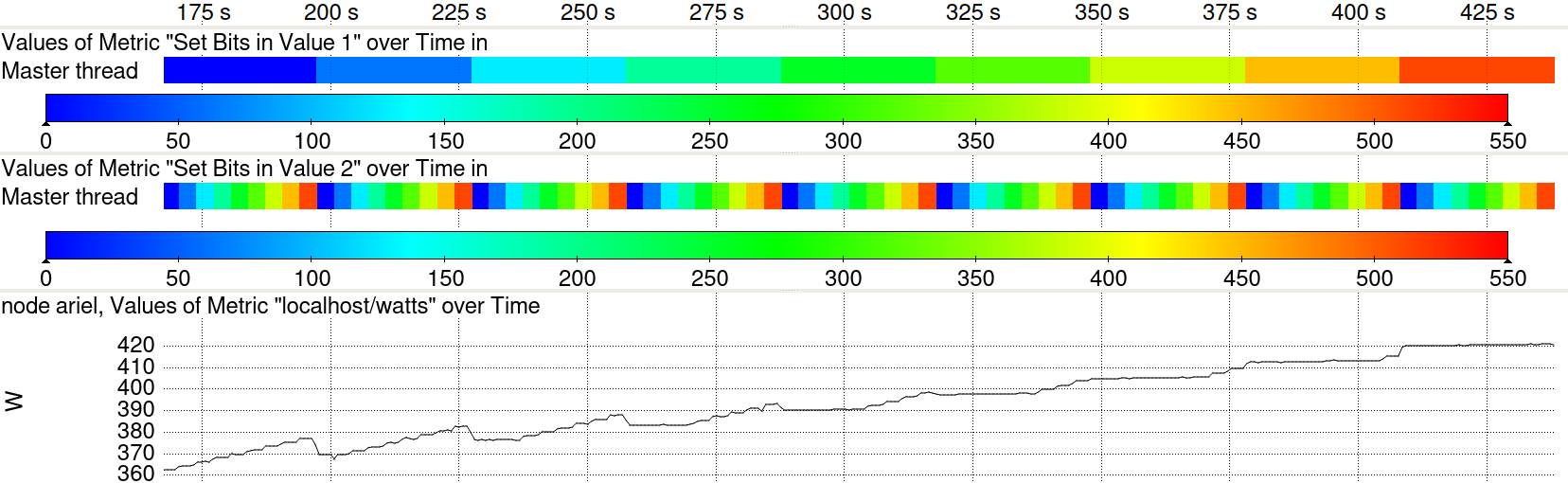}
	\caption{\label{fig:power-var-trace}Vampir trace visualization of data-dependent power at nominal frequency.
		Over time, the bit count of $value_1$ and $value_2$ is changed (top and middle panel).
		Full system AC power consumption shown is in the bottom panel.}
\end{figure*}

\setcounter{figure}{6}
\begin{figure}[t]
\center
	\includegraphics[width=.75\columnwidth]{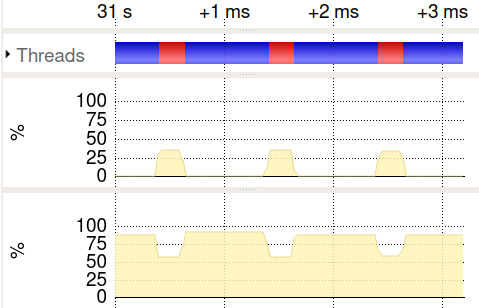}
	\caption{\label{fig:AVX-measurement-vampir-wc} Vampir visualization of worst-case scenario of AVX frequency changes - alternating between \textit{High} (200\,µs) and \textit{Low} (800\,µs) phases (top panel, red and blue, resp.).
	When switching to AVX frequency, the out-of-order engine is throttled.
	Here, more than 30\,\% of the cycles are spent throttled (middle panel).
	In the \textit{Low} phase, more than 85\,\% of the cycles are spent at the reduced AVX-512 frequency range even though no AVX instruction is executed (bottom panel).}
\end{figure}
Initially, we use the nominal frequency of the processor and run one thread per core with a period of 2\,s (50\,\% \textit{Low}) for five minutes (150 iterations)\footnote{\texttt{taskset -c 0-35 ./FIRESTARTER -l 50 -p 2000000 -t 300}}. %
The measurements show that the total number of cycles spent with a throttled OoO-engine varies between the threads, ranging between 28 and 34 million.
This translates to 62\,µs and 75\,µs for each of the 150 iterations.

Also the number of cycles that are spent in AVX-512 frequency during \textit{Low} varies between 1.5 and 1.9 million (555\,µs to 704\,µs at AVX-512 frequency of 2.7\,GHz).
With a reduced period of 2\,ms, the band narrows down to 1.8 to 1.9 million cycles of AVX-frequency range during each \textit{Low} phase.
With an unbalanced workload (e.g., 1\,ms period with 80\,\% of the time being spent in \textit{Low}), a worst-case scenario can be created (see \figref{fig:AVX-measurement-vampir-wc}).
A similar situation can occur if an application calls a highly optimized mathematical routine in a loop, while doing additional non-AVX instructions between these calls.

This shows, that the AVX frequency mechanisms can have a significant and non-obvious influence on the performance of applications, in particular during transitions.
Throttling of AVX instructions and a latency of several hundred microseconds for restoring non-AVX frequency need to be considered.
To avoid out-of-order throttling, we reduced the frequency of the system to 1.2\,GHz, which is lower than any documented AVX frequency.
Even in this situation, the instruction issue is throttled according to the PMC measurements.
However, the measured OoO-throttling cycles have a lower variance compared to measurements at higher core frequencies.

\section{Data-Dependent Power Consumption}
\label{sec:power}
Skylake-SP processors are the first x86 standard processors, besides the Xeon Phi, to support 512 bit wide vector operations.
Together with the increased number of cores, this leads to a significantly increased number of bits that can change in a processor within a single cycle. %
A possible consequence is that power consumption may vary depending on the input \emph{data}, even with a fixed set of instructions.
To test this, we designed an OpenMP-parallel benchmark, which only uses data in 512-bit-wide registers, and execute it on all cores.
The benchmark places $value_1$ in \texttt{zmm[0-7]} and $value_2$ in \texttt{zmm[8-15]} and executes $value_2=value_2 \oplus value_1$ in a loop for several seconds.
This forces the processor to use the AVX frequency, since \texttt{vxorps} can be considered a ``light'' AVX-512 instruction.
Moreover, we measured that core frequency and the instruction rate are constant during the experiment.
We repeated the workload for multiple frequencies while collecting at least 50 power samples for each frequency/$value_1$/$value_2$ combination.

In order to visualize the effect, we generated traces using Score-P with power measurements.
\figref{fig:power-var-trace} shows that the power consumption strongly correlates with the used data.
Only depending on the number of set bits in $value_1$ and $value_2$, the power consumption ranges from 362\,W to 420\,W for a 3\,GHz setup.

For the further analysis, we use the mean of the measurements, while disregarding the first 10 and last 5 power measurements, as they may contain transitions.
The population count of $value_1$ defines how many bits in the target registers \texttt{zmmm[8-15]} will toggle with every instruction, independent of the setting of $value_2$.
The power consumption increase correlates with that count.
Furthermore, the power consumption increases for each bit set in $value_2$ if $popcnt(value_2) > popcnt(value_1)$.
Here, the additional bits ($popcnt(value_2) - popcnt(value_1)$) within the register will not change with the operation, but will be read as, and written to be, one instead of zero.
We use linear regression to determine the power consumption increase per bit set of $value_1$ per core.
For 2.4\,GHz and 3.0\,GHz, the power consumption of the system increases by 1.69\,mW and 3.13\,mW per bit set in $value_1$, respectively.
The costs for each additional bit in $value_2$ are 0.46\,mW and 0.80\,mW for 2.4\,GHz and 3.0\,GHz, respectively.

When comparing the share of data-dependent power costs against the base system power consumption, the shape of the used data can cause a difference of up to 78\,W at turbo frequency.
If we subtract a base power consumption for active cores and uncore components (C1 state), the power costs of the single instruction increases by more than 30\,\% with all bits being set, independently of the frequency.
This far-reaching implication of this insight is that any accurate instruction-based power and energy model for this processor needs to consider input data.

\section{Conclusion and Future Work}
\label{sec:summary}
In this paper, we have described details of the Skylake-SP energy efficiency mechanisms and how they influence performance and power consumption.
We demonstrated that neither the P-state nor the T-state mechanism has been improved over the Haswell-EP platform.
Furthermore, the C-state latencies have increased.
The uncore frequency also has a significant influence on performance, as the default mechanism takes about 10 milliseconds to adapt to application behavior changes.
Another transparent mechanism, AVX-512 frequencies, will stall the out-of-order engine for about 70\,µs after the first AVX-512 instruction has been issued.
Moreover, AVX-512 frequencies will be applied more than 600\,µs after the last instruction has been issued.
This may lead to pathological cases in which the processor is constantly switching between two non-optimal configurations.
Another effect concerns the wider AVX-512 registers: the dynamic part of the power consumption increased significantly.
We demonstrated that--depending on the input data--the power consumption of a workload can differ by more than 18\,\% for the whole system (or more than 30\,\% if only the dynamic power on top of C1 is considered). %
This is a major challenge and may significantly limit the usefulness of any instruction-based energy model that is unaware of the input data.
Our future work will consider the analysis of competitor systems and the influence of energy efficiency features on latencies and bandwidths.
Furthermore, we will investigate the processor energy counters more closely and extend the system stress test FIRESTARTER.

\section*{Acknowledgments}

This work is supported in part by the European Union’s Horizon 2020 program in the READEX project (grant agreement number 671657) and by the German Research Foundation (DFG) within the CRC 912 - HAEC.

\section*{Used Code}
The used code is available at \url{https://github.com/tud-zih-energy/2019-HPCS-Skylake-EE}.

\bibliographystyle{IEEEtran}
\bibliography{paper}

\end{document}